\documentclass[aps,twocolumn,preprintnumbers,showpacs]{revtex4}

\usepackage{epsfig}
\usepackage{amsmath,amssymb}
\usepackage{graphicx}
\usepackage{revsymb}

\newcommand{\be}{\begin{equation}}
\newcommand{\ee}{\end{equation}}
\newcommand{\bq}{\begin{eqnarray}}
\newcommand{\eq}{\end{eqnarray}}

\def\H{{\cal H}}

\begin{document}

\title{Phenomenological implications of an alternative Hamiltonian constraint for quantum cosmology}
\author{Mikhail Kagan \footnote{e-mail address: {\tt mak411@psu.edu}}}
\vspace{.5cm} \affiliation{Department of Physics, Pennsylvania State
University, University Park, PA 16802, USA} 
\begin{abstract}
In this paper we review a model based on loop quantum cosmology
that arises from a symmetry reduction of the self dual Plebanski
action. In this formulation the symmetry reduction leads to a very
simple Hamiltonian constraint that can be quantized explicitly in
the framework of loop quantum cosmology. We investigate the
phenomenological implications of this model in the semi-classical
regime and compare those with the known results of the standard
Loop Quantum Cosmology.
\end{abstract}

\maketitle

\section{Introduction}

Modern observations indicate that, at least on large scales, the
Universe appears to be homogeneous and isotropic. In the framework
of General Relativity, such a universe is best
described by the Friedmann-Robertson-Walker model, which nowadays
is the accepted paradigm of cosmology. While the full
(non-homogeneous and non-isotropic) theory has infinite degrees of
freedom, cosmological models are characterized by a
finite-dimensional phase-space. The latter makes cosmology a
fruitful testing arena for existing theories of gravity.

One of the central issues of general relativity is the development
of space-time singularities. One famous instance is the occurrence
of the initial Big-Bang singularity in cosmological models, which
takes place under quite general assumptions. As one approaches the
Big-Bang singularity, space-time curvature and energy/mass
densities blow up, and any description based on classical notion
of space-time fails. One is forced to take quantum effects into
account and replace General Relativity by a quantum theory of
gravity.

Aside from the initial Big-Bang singularity stands the issue of
reconciling modern observations with predictions based on the
Friedmann equations. Among others there are the {\em flatness} and
{\em horizon} problems. These have been successfully explained by
the mechanism called {\em inflation}, by introducing matter in
theory in the form of a massive scalar field $\phi$, {\em
inflaton}, minimally coupled to gravity \cite{LL}. According to
the inflationary paradigm, there has to have been an epoch, when
the Universe was undergoing an accelerated expansion; such that
the size of the Universe increased by a factor of about $e^{60}$
(60 e-foldings) to agree with observations. The mechanism,
proposed in \cite{LL}, considers so called {\em slow-roll
approximation}, when the inflaton rolls down its potential hill
towards the potential minimum, but, at the same time, remaining
far away from the minimum. It is exactly in the vicinity of the
minimum of the potential, where the inflation stops. For the
simplest choice of a quadratic field potential with an arbitrary
mass of the inflaton, one would obtain 60 e-foldings if the
slow-roll regime starts of at $\phi \approx 3 M_{\rm P}$ and ends
when the field is essentially zero. The question as to how the
inflaton gets to that value before the slow-roll is left for a
quantum theory to answer.

Loop quantum gravity (LQG) is believed to be the most plausible
candidate for a theory which would reconcile General Relativity and
Quantum Mechanics \cite{LQG,ap}. It is a background independent non
perturbative approach to the canonical quantization of general
relativity. The prime prediction of the theory is the discreteness
of geometry at the fundamental level: operators corresponding to
geometric quantities such as the area and volume have discrete
spectra \cite{geometrical_op}. Geometry gets quantized and at
Planckian scales the smooth notion of space and time ceases to
exist.

Loop quantum gravity, restricted to a mini-superspace quantization,
is called loop quantum cosmology (LQC) \cite{ICGC}. The major
simplifications due to the symmetry reduction make it possible to
address physical questions that still remain open in the full
theory. For instance LQC provides means for the understanding and
resolution of the Big-Bang singularity. The important physical
question of whether classical singularities of general relativity
are absent in the quantum theory is answered in the affirmative in
LQC \cite{Bojowald:2001xe,ABL,sing-res} (for a recent review see
\cite{ICGC}). This remarkable result originates from the fundamental
discreteness of geometric operators, which dramatically changes the
quantum `evolution' equation in the deep Planckian regime. It has
also been shown that LQC leads to modifications of the classical
Friedmann equations \cite{ModFried,2ndCorr}. Recent works have shown
that these effects can provide the proper initial conditions for
chaotic inflation \cite{Ambiguities,ChaInf,closedinflation} and may
lead to measurable effects on the CMB \cite{CMB}.

LQC studies the quantum evolution of a homogeneous and isotropic
universe. One starts from the symmetric Einstein-Hilbert action and
performs the Legendre transform to obtain the Hamiltonian
constraint. The wave function of the universe is to be annihilated
by the quantum counterpart of the constraint. By virtue of the
underlying discreteness, the equations  governing the evolution in
the deep Planckian regime are difference equations rather than
differential equations. In other words, as the notion of the
infinitesimal translation ceases to exist at this scale, only an
operator of a finite translation can be well-defined, thus replacing
the common derivation operator.

It is, however, sensible to expect that at some scale, $a_0$, the
space-time would become essentially smooth and continuous. Indeed,
it was shown in \cite{Semi,Semi_SV} that in the continuum limit the
difference equation yields the well-known Wheeler-deWitt equation,
which is an indication in favour of the viability of LQC. On the
other hand, it would make sense to suppose that there exist an
intermediate range of scale factors, $a_0 \lesssim a \lesssim a_*$,
for which the space time is not yet classical, i.e. the
non-perturbative modifications play a substantial role. This
approximation is called {\em semi-classical}.

To make the notion of semi-classicality more precise, we should
specify the upper and lower bounds of the scale factor. In
\cite{Semi,Semi_SV} shown that the difference equations are well
approximated by Friedmann differential equations for scale factors
above $a_0\approx \sqrt{\gamma \mu_0}l_{\rm P}$ ($\mu_0 \equiv
\sqrt{3}/4$ being quantum ambiguity parameter related to the lowest
eigenvalue of the area operator in LQG \cite{ABL}). Here $l_{\rm
P}=\sqrt {G \hbar}$ is the Planck length and $\gamma\approx 0.2375$
is the Barbero-Immirzi parameter, controlling discreteness of the
quantum space-time, whose value is determined from black hole
thermodynamics \cite{bek_hawking1}. For a minimally coupled scalar
field, the quantum corrections manifest themselves via the
geometrical density (inverse volume) operator. The crucial result of
LQC is that the spectrum of this operator turns out to be bounded
from above (classically it blows up) as $a \rightarrow 0$. Moreover,
for small scale factors it becomes proportional to a positive power
of $a$. The classical behaviour of the geometrical density operator
is recovered soon after $a\geq a_*\equiv \sqrt{\gamma \mu_0
j/3}l_{\rm P}$ ($j$ being a half-integer, greater that unity), thus
defining the upper bound of the semi-classical regime.

One of the open issues of LQG that remains is the characterization
of the physical Hilbert space, which needs a physical inner
product and a notion of unitary evolution. That, in turn, requires
one to construct a self-adjoint Hamiltonian constraint operator.
As the standard Hamiltonian obtained from the Einstein-Hilbert
action is not self-adjoint, one has to symmetrize it in a
non-trivial way \cite{Semi, Willis}. An alternative way, usually
taken by Spin Foams theorists, is based on the Plebanski action
\cite{pleb}, which classically is equivalent to the
Einstein-Hilbert action. In this formulation the symmetry
reduction leads to a very simple Hamiltonian constraint that is
self-adjoint and can be quantized explicitly in the framework of
LQC \cite{NPV}. The model is defined in the Riemannian sector,
which constitutes a limitation when considering physical
predictions. However, a method for transforming to the Lorentzian
sector was shown in Ref.~\cite{NPV}, and it was argued that the
main results of the model hold true in the Lorentzian sector.

In this paper we review a model based on LQC that arises from a
symmetry reduction of the self-dual Plebanski action. We show that
in the homogeneous and isotropic case, the Plebanski action can be
viewed as the standard one (symmetry reduced Einstein-Hilbert
action) with a specific choice of the lapse function. As far as
the lapse plays the role of a Lagrange multiplier imposing the
Hamiltonian constraint, both actions must agree classically and
yield the same physics. Upon quantization, however, the
arbitrariness of lapse gives rise to yet one more (lapse)
ambiguity. We must, nevertheless, emphasize that the two actions
cannot be related merely by a choice of lapse in the full theory,
the correspondence holds only for the symmetry-reduced models.
Therefore, from now on we will avoid using the word `ambiguity',
as it is commonly referred to in the context of the full theory.
Various quantization ambiguities present in LQC are discussed in
great detail in \cite{Ambiguities}. In current paper we
deliberately leave out all ambiguities (except for the parameter
$j$) and focus on the lapse arbitrariness. (In particular, we fix
the parameter $l$ appearing in the geometrical density operator
(see \cite{Ambiguities}) at its natural value $l=3/4$.)

While this specific choice of lapse appears automatically, when
obtained as the result of the classical symmetry reduction of the
full theory, its explicit form is, indeed, natural, as it leads to
a self-adjoint Hamiltonian constraint.

Generally, such a lapse arbitrariness would lead to different
quantum theories in both deep Planckian regime and semi-classical
regime. We are focusing on the semi-classical approximation. The
goal of this work is to investigate the robustness of the theory
with respect to this kind of `ambiguity'. We show that the
quantum-corrected Friedmann equations indeed lead to
(semi-classical) inflation, which provides conditions for the
classical slow-roll regime. Also, the solutions for both standard
and Plebanski models may differ substantially during the
semiclassical inflation, but get very close to each other in the
classical domain. Moreover, for observationally interesting
solutions (those corresponding to the maximum value of the inflaton
of about $3{\rm M}_P$) the difference between the two models is
insignificant even within the domain of the semi-classical
approximation.

\section{Classical Theory}

%In this section we revise the standard formulation of LQC starting
%with the action of the full theory. We then realize symmetry
%reduction and proceed to homogeneous and isotropic model.

In this section we illustrate the basic ideas underlying LQC.
Conventionally, one starts with the Einstein-Hilbert action of
General relativity and performs its symmetry reduction, assuming
isotropy and homogeneity. The simplified action, written in the
Hamiltonian formulation, is subject to the standard Dirac
quantization. It turns out that the Hamiltonian of the system is
weakly zero, and the dynamics is governed by the (Hamiltonian)
constraint.

Firstly, we review the standard method used to obtain the
Hamiltonian constraint from the Einstein-Hilbert action in the
framework of the FRW model and derive Friedmann's equations.
Secondly, in section (\ref{std}) we consider an alternative
formulation of the gravitational action, which yields equivalent
classical equations of motion, but proves to be different upon
quantization.

\subsection{Standard formulation} \label{std}
The Einstein-Hilbert action is given by
\be \label{EH} S_{EH}=\frac{1}{2\kappa} \int d^4x \sqrt{|g|}R ,
\qquad \qquad (\kappa=8 \pi G) \ee
($g$ being the determinant of the metric and $R$ the Ricci scalar)
and consider a general flat and isotropic model
\be g_{ab}dx^adx^b=-N^2dt^2+a^2d{\vec x}^2 \ee
Here $N$ is the lapse function and $a$ is the scale factor.
Calculating the scalar curvature as usual, and inserting into
(\ref{EH}), we obtain
\be S_{EH,\rm iso}=\frac{1}{2\kappa} \int \frac{dt}{N} \left[6a^2
\ddot a + 6a \dot a^2 \right] \ee
We then proceed to introduce the {\it connection} $A_a^{i}$ and
densitized {\it triad} $\tilde E_i^a$ variables as follows:
\bq
\label{def} \tilde E_i^a \tilde E_i^b&=&|q|q^{ab}\\
\label{def2} A_a^i&=&\Gamma_a^i+\gamma K^i_a \eq
where $q_{ab}$ represents the spatial metric, $\Gamma_a^i$ the
intrinsic curvature, $K_a^i$ the extrinsic curvature and $\gamma$
the Barbero-Immirzi parameter. Due to flatness and isotropy, these
quantities can be expressed in terms of two (time-dependent)
numbers as follows:
\be \label{iso} \tilde E_i^a=c \delta_i^a, \qquad A_a^i=p \delta_a^i
\ee
Using the definitions (\ref{def}) and (\ref{def2}) and the fact
that in a flat isotropic spacetime we have:
\be \Gamma_a^i=0, \qquad K_a^i=\frac{\dot a}{N} \delta_a^i \ee
we obtain an expression for $p$ and $c$ in terms of $a$, $\dot a$:
\be p=a^2, \qquad c=\gamma \frac{\dot a}{N} \ee
from which:
\be \label{S_LQC} S_{LQC}=\frac{3}{\kappa} \int dt \left[\frac{p
\dot c}{\gamma}+N\frac{\sqrt{p}c^2}{\gamma^2}\right] \ee
Note that one can, of course, arrive at the same expression
inserting $A_a^{i}$ and $\tilde E_i^a$ of (\ref{iso}) into the
action of the full theory \cite{ABL}:
\bq H_{\rm gr} &:=& \int d^3x N\, e^{-1} \left( \epsilon_{ijk}
F_{ab}^i \tilde E^{aj} \tilde E^{bk} \nonumber \right. \\
 &-& 2(1+\gamma^2)K_{[a}^iK_{b]}^j \tilde E_i^a \tilde E_j^b \left. \right) \nonumber\\
&=& -\gamma^{-2}\int d^3x\,  N\, \epsilon_{ijk}F_{ab}^i\,\, e^{-1}
{\tilde E^{aj} \tilde E^{bk}} \eq
where $e:=\sqrt{|\det E|}\,\, \text{sgn}(\det E)$, $F_{ab}^i$ is the
curvature of the connection. In the second step we have used the
fact that for spatially flat, homogeneous models the two terms in
the full constraint are proportional to each other. Technically the
integral diverges, since the left invariant metric is constant and
the integral is over a non compact manifold. To overcome this, we
restrict the integral to a finite cell with volume $V_0$ and absorb
this factor into the variables as $A_a^{i} \, (p) \rightarrow
V_0^{2/3} A_a^{i} \, (p)$ and $\tilde E_i^a \, (c) \rightarrow
V_0^{1/3} \tilde E_i^a \, (c)$. By doing so, one obtains the action
(\ref{S_LQC}).

The action (\ref {S_LQC}) is expressed in terms of the canonical
variable $c$ and its conjugate momentum $p'=3p/ \kappa \gamma$.
This yields the Poisson brackets: \be \{c,p\}=\frac{\kappa
\gamma}{3}. \ee
\\
Because of homogeneity, the lapse $N$ can depend only on time and
plays the role of a Lagrange multiplier: $N$ being freely
specifiable, the coefficient of $N$ in the action needs to be
identically zero, yielding the Hamiltonian Constraint
\be \label{HG} \H_{\rm gr} \equiv -\frac{3 \sqrt{p} c^2}{\kappa
\gamma^2}=0 \ee
or, if matter is present in the form of a scalar field $\phi$,
with Hamiltonian
\be \label{HM} \H_{\phi}=\frac{p_\phi^2}{2 p^{3/2}}+p^{3/2}
V(\phi), \ee
$p_\phi$ being the field momentum, then
\be \label{H_std} \H=-\frac{3 \sqrt{p} c^2}{\kappa
\gamma^2}+\H_{\phi}=0 \ee
The Hamilton equations of motion now follow: \bq
 \dot p&=&\{p,N\H\}=-\frac{\gamma\kappa}{3} \frac{\partial \H}{\partial c} = \frac{2c}{\gamma}\sqrt{p} \\
 \dot c&=&\{c,N\H\}=\frac{\gamma\kappa}{3} \frac{\partial
\H}{\partial p} \nonumber \\
&=& \frac{\gamma\kappa}{2}\left(\frac{c^2}{2\gamma\sqrt{p}}-\frac{p_\phi^2}{p^{5/2}}+\sqrt{p} V(\phi)\right)\\
 \dot \phi&=&\{\phi,N\H\}=\frac{p_\phi}{p^{3/2}} \\
 \dot p_\phi&=&\{p_\phi,N\H\}=-p^{3/2} V_{,\phi}(\phi) \eq
We have also used the lapse freedom and set $N=1$. Note that only
three phase variables are independent. For later convenience, we
shall eliminate one of them (specifically, the connection $c$) using
the Hamiltonian constraint (\ref{H_std}). One can easily show that
the equations above lead to familiar Friedmann's equations. The
constraint equation, rewritten in terms of the scale factor and its
time derivative, reads
\be \H=-\frac{3}{\kappa}a \dot a^2+\frac{a^3}{2}\dot \phi^2+a^3
V(\phi) \approx 0 \ee Dividing by $a^3$, one gets the first
Friedmann equation:
\be \label{Fri} H^2=\frac{\kappa}{3}\left[\frac {\dot
\phi^2}{2}+V(\phi)\right] \ee
Here $H \equiv \dot a/a$ is the Hubble rate. Using the $\dot
p_\phi$-equation we derive the Klein-Gordon equation
\be \label{KG} \ddot \phi+3 H\dot \phi+V,_\phi(\phi)=0 \ee
Finally, the Raychaudhuri equation can be obtained by taking the
time derivative of (\ref{Fri}) and substituting $\ddot \phi$ from
(\ref{KG}):
\be \label{Ray} \dot H=-\frac{\kappa}{2}\dot \phi^2 \ee
These equations particularly tell us, that for any reasonable
scalar field and an expanding universe ($H>0$), (\ref{KG}) looks
like an oscillator equation with dissipation, whereas (\ref{Ray})
implies decelerated expansion. We shall see that if quantum
effects are taken into account, the universe can expand
accelerating (super-inflation), and the friction term can become
negative, thus driving the scalar field (inflaton) up-hill.

\subsection{Plebanski formulation}
We start, following \cite{NPV}, by considering the classical
self-dual Riemannian gravitational action, symmetry reduce it to
the homogeneous and isotropic model and show that the system would
be governed by equations of motions equivalent to the standard
ones of the previous section.

As noted by Plebanski\cite{pleb} the classical self-dual
Riemannian gravitational action can be written as
\begin{equation}
\label{full_BF_action}
    S[A,\Sigma, \psi] = \frac{2}{\kappa}\int  \Sigma_i \wedge F^i(A)  -
    \psi_{ij} \left[ \Sigma^i \wedge \Sigma^j -
        \frac{1}{3} \delta^{ij} \Sigma^k \wedge \Sigma_k \right]
\end{equation}
where $\Sigma=\Sigma^i_{\mu \nu} dx^{\mu} dx^{\nu} \tau_i$ is an
SU(2) Lie algebra valued two form, $A=A^i_{\mu} dx^{\mu} \tau_i$
is an SU(2) Lie algebra valued connection, $\tau_i$ are the
generators of the Lie algebra of SU(2). The tensor $\psi_{ij}$ is
symmetric and acts as a Lagrange multiplier enforcing the
constraint $ \Sigma^i \wedge \Sigma^j - \frac{1}{3} \delta^{ij}
\Sigma^k \wedge \Sigma_k = 0$. Once this constraint is solved, the
action becomes equivalent to the self-dual action of general
relativity. The model is defined in the Riemannian sector which
constitutes a limitation when considering physical predictions.
However, in \cite{NPV} was shown a method for transforming to the
Lorentzian sector and argued that the main results of the model
are manifested in the Lorentzian sector.

In order to reduce the action (\ref{full_BF_action}) to spatial
homogeneity and isotropy, we write the action explicitly in terms
of coordinates separating space and time (which we here call
$\tau$ to avoid confusion with the previous section) as \bq
\label{split_BF_action}
    S[A,\tilde E, B] &=&  \frac{2}{\kappa}\int d\tau \int d^3\!x
     A^i_{a,\tau} \tilde{E}^a_i+A^i_0 D_a \tilde{E}^a_i
        +\epsilon^{abc} B^i_a F^i_{bc}  \nonumber \\
    &-& \psi_{ij}  \left[\frac{1}{2}  B^i_a \tilde{E}^{aj} +
        \frac{1}{2} B^j_a \tilde{E}^{ai} -
    \frac{4}{3} \delta^{ij} B^k_a \tilde{E}^a_k \right]
\eq where we introduced the definitions $\tilde E^{ai} (\equiv
\sqrt{q} E^{ai}) \equiv  \, \epsilon^{abc} \Sigma^i_{bc}$, $B^i_a
\equiv 2 \, \Sigma^i_{0a}$, and $\epsilon^{abc}=\epsilon^{0abc}$.
Assuming spatial homogeneity and isotropy, we take the triad and
connection of the form (\ref {iso}) and, similarly,
$B_a^i=B\delta_a^i$. Written in terms of the reduced variables the
homogeneous and isotropic gravitational action becomes
\begin{eqnarray}\label{isotropic_action}
    S_{\rm
Pleb}[c,p,B] &=& \frac {3}{\kappa}\int\!\! d \tau \int\!\!  d^3\!x
    \; (\sqrt{q}  p c_{,\tau} + \sqrt{q}  B c^2)  \nonumber \\
    &=& V_0 \frac{3}{\kappa}\int\!\! d\tau \; (p c_{,\tau} + B c^2)
\end{eqnarray}
where $V_0 \equiv \int \!\!  d^3\!x \sqrt{q}$. The constraint term
$ \psi_{ij} \left[ B^i_a \tilde{E}^{aj} + B^j_a \tilde{E}^{ai} -
        \frac{1}{3} \delta^{ij} B^k_a \tilde{E}^a_k \right]$
in the full action (\ref{split_BF_action}) vanishes identically
for the isotropic model which can be checked by a direct
calculation. Here we again integrate over a finite cell with volume $V_0$ and
absorb this factor into the variables as $p \rightarrow
V_0^{2/3}p$, $c \rightarrow V_0^{1/3} c$, and $B \rightarrow
V_0^{1/3} B$, whence the action becomes
\bq\label{S_Pleb} S_{\rm Pleb}(c,p,B) = \frac {3}{\kappa}\int\!\!
d\tau \;  (p c_{,\tau} + B\, c^2)  . \eq
Since $\H_{\rm gr} \approx 0$ (see \ref{HG}), the action
(\ref{S_LQC}) admits time reparametrization. In particular,
setting $d\tau$=$\sqrt{p}dt$, one obtains the action above. Taking
this into account, one can formally introduce the matter part in
the action as:
\be \label{HPl} \H_{\rm
Pleb}=-\frac{3c^2}{\kappa\gamma^2}+\frac{1}{\sqrt{p}}\H_{\phi} \ee
Note that $\H_{\rm Pleb}$ does not have units of energy while
$\H_\phi$ does. Hence we have written the prefactor
$\frac{1}{\sqrt{p}}$, such that $\H_\phi$ coincides with the
standard form of the matter Hamiltonian (Eq. \ref{HM}). We
conclude the section with Hamilton's equations of motion
\bq \label{Heq_Pl_Classsic}
&& p_{\tau}=\{p,B\H_{\rm Pleb}\}=\frac{2c}{\gamma} \\
&& c_{\tau}=\{c,B\H_{\rm Pleb}\}=\frac{2\gamma\kappa}{3}\left(-\frac{p_\phi^2}{p^3}+V(\phi)\right)\\
&& \phi_{\tau}=\{\phi,B\H_{\rm Pleb}\}=\frac{p_\phi}{p^2} \\
&& p_{\phi,\tau}=\{p_\phi,B\H_{\rm Pleb}\}=-p V_{,\phi}(\phi) \eq
Here we similarly set $B=1$ to arrive at the righthand side of the
equations of motion. From now on, we shall no longer use the
connection, $c$. We will rather express it in terms of the three
other phase variables, using Eq. (\ref{H_std})
\be \frac{c}{\gamma}=\sqrt{\frac{\kappa}{3}} \sqrt{p
V(\phi)+\frac{p_\phi^2}{2 p^2}} \ee
Note that Eq. (\ref{HPl}) yields the same result. It is also
straightforward to show that the Hamilton equations above
reproduce the standard ones, after the backward time
reparametrization $d\tau$=$\sqrt{p}dt$. One can also put it the
following way. Identify the time variables, $t$ and $\tau$, but
think of the two actions, (\ref{S_LQC}) and (\ref{S_Pleb}), as
described by different lapse functions. Specifically, the former
corresponds to $N=1$, whereas the latter is related to $B \equiv
\sqrt{p} N=1$. In other words, the two descriptions lead to the
same classical equations of motion, if one sets
\be \label{lapse} N=1/\sqrt{p}\ee
We shall,however, see that the loop quantization of the two
constraints (\ref{H_std}) and (\ref{HPl}) gives rise to a
discrepancy between the two models, and the quantum corrected
dynamics shall be different in the semi-classical approximation.

\section{Quantization}
In order to {\em quantize} a theory, one promotes mutually conjugate
variables to operators, whose commutation relation derives from the
Poisson bracket of the classical theory. In the framework of the
conventional (Schr{\"o}dinger) quantization, the operators are
defined in a way that one of them acts as a multiplication, while
the other one is a derivation. Applied to cosmology this leads to
the famous Wheeler-DeWitt equation which provides an accurate
description of the wavefunction of the universe at relatively large
scales, but fails in describing classical singularities. For at the
Planckian scale smooth space-time is no longer a meaningful notion,
hence derivative cannot be well-defined. The problem can be solved
in the framework of another quantization scheme, which is based on
holonomies around closed loops and is called {\em Loop
Quantization}. The basic idea is that one needs to replace
derivatives or, more precisely, differentials (that can be thought
of as infinitesimal translations) with (finite) translations
respecting the fundamental discreteness of space-time in the deep
Planckian regime. In this section we consider the Loop Quantization
procedure for the constraints (\ref{H_std}) and (\ref{HPl}) and then
discuss the implications in the semi-classical regime.

\subsection{Standard formulation}
The lesson taught by the full theory states that if one works in
the tetrad representation, the connection cannot be promoted to a
well defined operator. Instead one is forced to reformulate the
theory in terms of {\em holonomies} of the connection, whose
operators constitute translations. The main result of the loop
quantization, that we would need in this paper, is the spectrum of
the geometrical density operator $(\widehat{1/a^3})$. If we
consider the matter Hamiltonian (\ref {HM}), we see that
(classically) the kinetic term apparently blows up as $p$
approaches zero. Moreover, the naive attempt to write the
geometrical density operator $(\widehat{1/a^3})$ as $\hat a^{-3}
\equiv \hat p^{-3/2}$ fails simply because the triad operator has
a zero eigenvalue in its discrete spectrum, hence cannot be
inverted. Nonetheless, the inverse scale factor operator {\em can}
be defined  in terms of a commutator of well behaved operators
\cite{ABL,closedinflation}. Classically we have
\be a^{-3}=\left[ \frac{1}{2 \pi G \gamma} \left\{c,
|p|^{3/4}\right\} \right]^6 \ee
which after quantization becomes \cite{ICGC}
 \be d_j(a) = D(q) \, a^{-3}, ~~ q := a^2/a_*^2,
~~ a_* := \sqrt{j \gamma \mu_0 /3} \, l_P \label{density} \ee
where
\bq \label{defD}
D(q) &=& q^{3/2} \left\{  \frac {8}{11} \right. \left[
(q+1)^{11/4}-|q-1|^{11/4}\right ] \nonumber \\ &-& \frac{8}{7} q
\left. \left[ (q+1)^{7/4}-{\rm sgn}(q-1) |q-1|^{7/4}\right ] \right
\}^6 \eq
The asymptotic behaviour of the density operator follows from the
equation above. Not only does $d_j$ remain bounded for $a
\rightarrow 0$, it becomes proportional to a positive power of the
scale factor. Specifically, $D(q) \propto a^{15}$ and $d \propto
a^{12}$, thus making the matter Hamiltonian (\ref{HM}) well behaved.
On the other hand, as $a$ goes to infinity, one  recovers the
classical dependence:  $D(q) \approx 1$ whereas $d \approx a^{-3}$.

We have seen that in the deep Planckian regime, when $a \approx
a_0 \equiv \sqrt {\gamma \mu_0} l_P$, one is forced to use the
difference equation. The classical description
(\ref{Fri},\ref{KG},\ref{Ray}) works for $a \gg a_*$. There is,
however, an intermediate range of scale factors \cite{Semi_SV},
$a_0 \ll a \ll a_*$, when space-time is essentially smooth, so
that differential equations can be used, and also quantum
corrections are significant, i.e. $D(q)$ behaves non-classically.
In that case, one would use the Hamilton equations dictated by the
Hamiltonian constraint, modified in the following way. Every
negative power of the scale factor should be replaced by the
appropriate power of $d_j$:
\be \label{HStd_SC} \H_{\rm Std} = -\frac{3 \sqrt{p} c^2}{\kappa
\gamma^2} + d_j \frac{p_\phi^2}{2}+p^{3/2} V(\phi), \ee
The quantum corrected Friedmann, Klein-Gordon and Raychaudhuri
equations would then read:
\bq \label{FriSC} && H^2=\frac{\kappa}{3}\left(\frac {\dot
\phi^2}{2}D(q)^{-1}+V(\phi)\right) \\
\label{KGSC} && \ddot \phi+3 H\dot \phi\left( 1-\frac{\cal D}{3} \right)+D(q) V,_\phi(\phi)=0 \\
\label{RaySC} && \dot H=-\frac{\kappa}{2}D(q)^{-1}\dot \phi^2
\left( 1-\frac{\cal D}{6}\right) \eq
where we have introduced the logarithmic derivative
\be \label{CalD} {\cal D} := \frac{d {\rm ln} (D)}{d {\rm ln(a)} }
\equiv \frac {\dot D}{H D}\ee
\subsection{Plebanski formulation}
Earlier we had pointed out, that one can pass directly to the
semi-classical regime starting from the homogeneous and isotropic
Hamiltonian constraint of the theory. The recipe was to replace
all classical negative powers of the scale factor with the
appropriate powers of the eigenvalues of the geometrical density
operator, $d_j(a)$ (\ref{defD}). The modified constraint becomes
\be \label{HPl_SC} \H_{\rm
Pleb}=-\frac{3c^2}{\kappa\gamma^2}+d_j^{4/3}\frac{p_\phi^2}{2}+p
V(\phi) \ee
here $c=\gamma a a_\tau$ as before, and
\be p_\phi=\phi_{,\tau} d_j^{4/3}\equiv \phi_{,\tau}D(q)^{-4/3}
a^4 \ee
Note that the major difference between the standard and Plebanski
effective theories arises from the $d_j^{4/3}$ factor of equation
(\ref{HPl_SC}). Rewriting the constraint as
\be \H_{\rm Pleb}=-\frac{3}{\kappa} a^2 a_{,\tau}^2 +
D(q)^{-4/3}a^4\frac{\phi_{,\tau}^2}{2}+a^2 V(\phi)=0\ee
we obtain the quantum corrected Friedmann, Klein-Gordon and
Raychaudhuri equations:
\bq \label{FriSC_Pl} && \tilde H^2=\frac{\kappa}{3}\left(\frac {
\phi_{,\tau}^2}{2}D(q)^{-4/3}+\frac{V(\phi)}{a^2}\right) \\
\label{KGSC_Pl} && \phi_{,\tau\tau}+4 \tilde H \phi_{,\tau}\left( 1-\frac{\cal D}{3} \right)+\frac{D(q)^{4/3}}{a^2} V,_\phi(\phi)=0 \\
\label{RaySC_Pl} && \tilde H_{,\tau}=-\frac{\kappa V(\phi)}{3
a^2}-\frac{2\kappa}{3}D(q)^{-4/3}\phi_{,\tau}^2 \left(
1-\frac{\cal D}{6}\right) \eq
note that the logarithmic derivative $\cal D$ defined in
(\ref{CalD}) is lapse independent, hence we keep the same
notation. Also, performing a direct calculation, one can show that
under $\tau \rightarrow t$ the equations above become very similar
to the ones of the previous section (\ref{FriSC}, \ref{KGSC},
\ref{RaySC}), up to the replacement $D \rightarrow D^{4/3}$ (and
hence ${\cal D} \rightarrow 4 {\cal D}/3$). Obviously, in the
classical limit: $D \approx 1$, ${\cal D} \approx 0$, hence the
two sets of equations coincide.

For numerical purposes, however, it is more convenient to use the
(quantum corrected) Hamilton equations of motion. As far as the
connection variable $c$ has been eliminated, they become a system
of three first order ordinary differential equations and thus can
be solved by standard means. We now restrict ourselves to the
quadratic field potential
\be V(\phi)=\frac {1}{2} m_\phi^2 \phi^2 \ee
 and write out the modified equations
\bq
&&\dot p =\sqrt{\frac{16 \pi}{3}}\sqrt{p^2\phi^2+D\left(\frac
{p}{\alpha}
\right)^{4/3}\frac{p_\phi^2}{p}} \\
&&\dot \phi = D\left(\frac{p}{\alpha}
\right)^{4/3}\frac{p_\phi}{p^{3/2}} \\
&&\dot p_\phi = -p^{3/2}\phi
\eq
where $\alpha \equiv \gamma j/3$, and the phase variables have
been written in the following units:
\be \label{units}
[p]=l_{\rm P}^2, \, [\phi]=l_{\rm P}^{-1} \equiv M_{\rm P}, \,
[t]=m_{\phi}^{-1}, \,[p_\phi]=m_\phi l_{\rm P}^2
\ee
Note, again, that the Hamilton equations for the standard LQC can
be recovered from the corresponding Plebanski ones by replacing
all the exponents (4/3) of $D(p/\alpha)$ with a unity. In the
following section we present some solutions for the equations
above for the both models and analyze how they compare with one
another.

\section{Phenomenological Implications}

In this section we introduce some typical solutions to the
Hamilton equations of motion. The key question is how one chooses
initial conditions. While the choice for $p_i\equiv a_0^2 = \gamma
\mu_0 (l_{\rm P}^2)$ seems rather obvious, the initial values of
the inflaton, $\phi_i$, and its conjugate momentum, $p_\phi$, is a
more subtle issue \cite{Ambiguities}.
% discussion of coherent states
Consider the inflaton as a wavepacket, localized around $\{\bar
\phi,\bar p_\phi\}$, with a spread \{$\Delta \phi,\Delta
p_{\phi}$\} constrained by the uncertainty principle (written in
the Planck units)
\be \label{UncPr} \Delta\phi \, \Delta p_\phi \ge 1 \ee
We are interested in the inflaton starting off near the potential
minimum. As the uncertainty principle does not restrict the
expectation values of the wavepacket, one can consistently specify
the initial values to be $\bar \phi=0=\bar p_\phi$. Then, from the
probabilistic point of view, the most likely values for the
inflaton are of order of the spreads $\Delta \phi$ and $\Delta
p_\phi$. Obviously the initial value $\phi_i$ should be much less
than $3M_{\rm P}$. Therefore we set $\phi_i=0.1(M_{\rm P})$, and
$p_{\phi i} \equiv 1/\phi_i = 10 (l_{\rm P})$. Note that had we
chosen a lower $\phi_i$, that would have led to a higher initial
momentum, thus making successful inflation more likely.

Let us now compare the initial energy of the inflaton field in the
Plebanski and standard models
\bq \label{Ham_init}
\H_{\rm Pleb}^{(\phi)}&=&D(q)^{4/3}\frac{p_\phi^2}{2p^{3/2}}+
p^{3/2} V(\phi)
\nonumber \\
\H_{\rm Std}^{(\phi)}&=&D(q) \frac{p_\phi^2}{2p^{3/2}}+ p^{3/2}
V(\phi)
\eq
Provided the same initial conditions for both models, compare the
corresponding values of the initial energy. Since $D(q)<1$ for
$a_0 \ll a_*$, we conclude that
\be H_{\rm Pleb}^{(\phi)}<H_{\rm Std}^{(\phi)} \nonumber \ee
The latter implies a lower inflation for the Plebanski model.

\begin{figure}
\begin{center}
\includegraphics[width=7.5cm,height=7.5cm, angle=270]{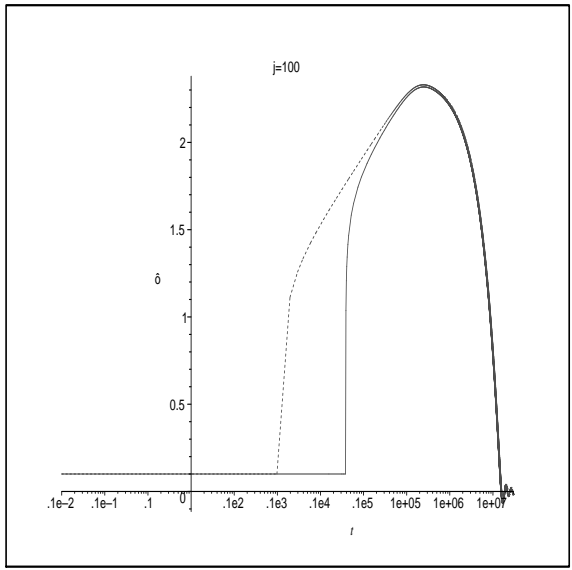}
\end{center}
\vskip-0.5cm \caption{Time dependence of the inflaton field for
$j~=~100$.
 Dashed and solid curves correspond to the standard and Plebanski models respectively} \label{fig1a}
\end{figure}

\begin{figure}
\begin{center}
\includegraphics[width=7.5cm,height=7.5cm, angle=270]{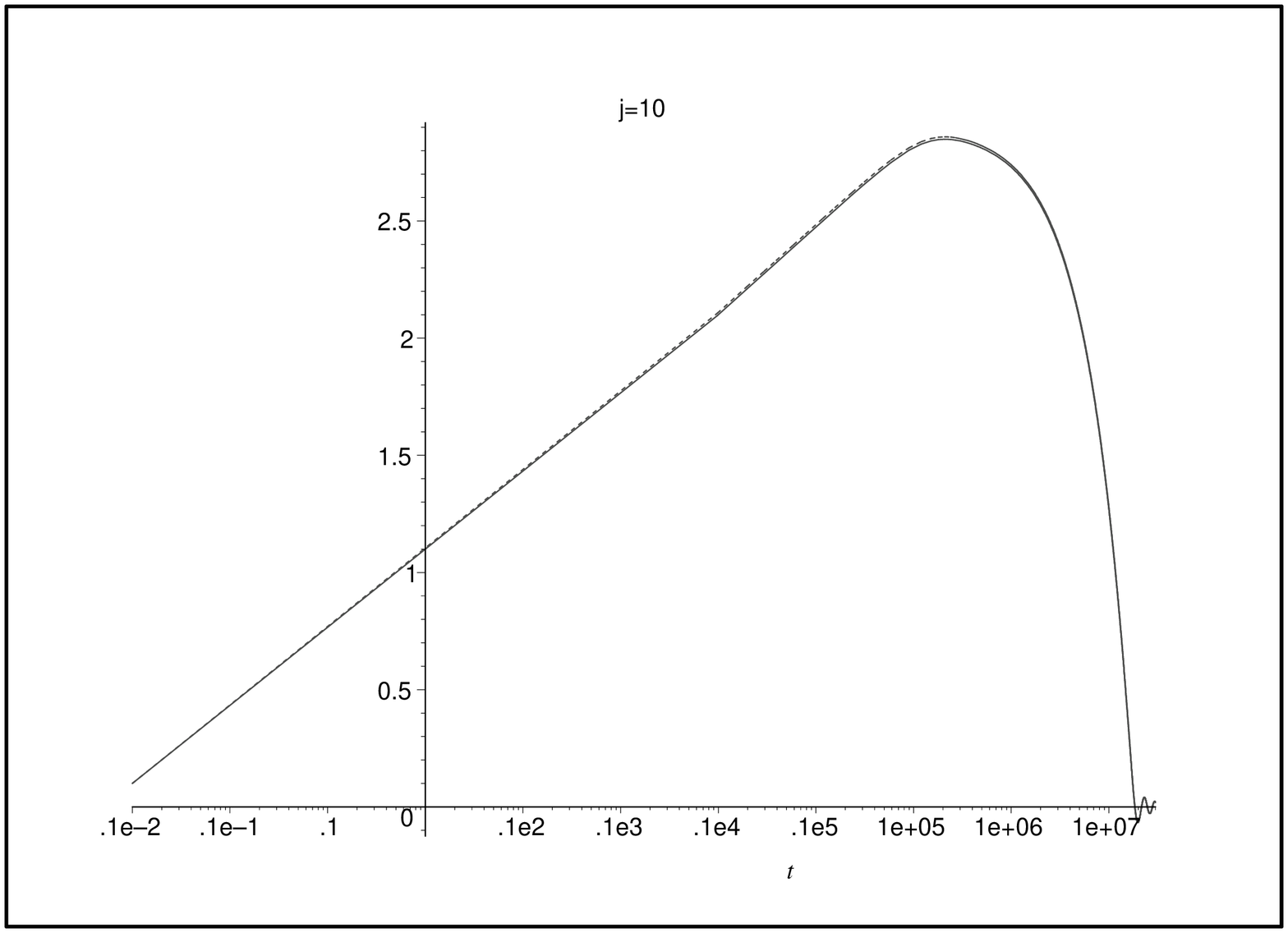}
\end{center}
\vskip-0.5cm \caption{Time dependence of the inflaton field for
$j~=~10$.
 Dashed and solid curves correspond to the standard and Plebanski models respectively} \label{fig1b}
\end{figure}

\begin{figure}
\begin{center}
\includegraphics[width=7.5cm,height=7.5cm, angle=270]{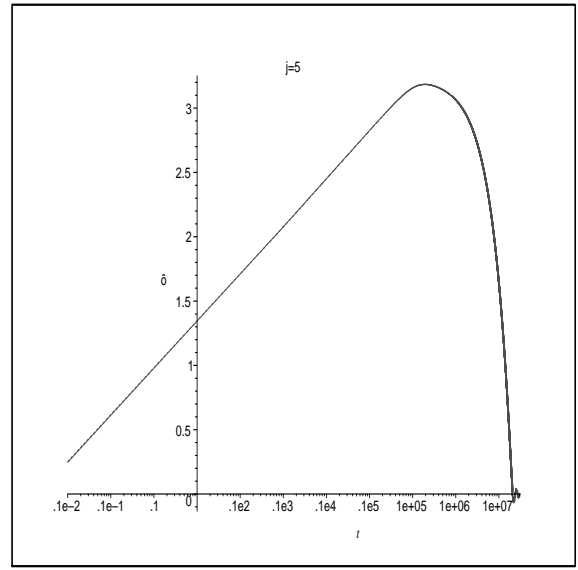}
\end{center}
\vskip-0.5cm \caption{Time dependence of the inflaton field for
$j~=~5$.
 Dashed and solid curves correspond to the standard and Plebanski models respectively} \label{fig1c}
\end{figure}

\begin{figure}
\begin{center}
\includegraphics[width=7.5cm,height=7.5cm, angle=270]{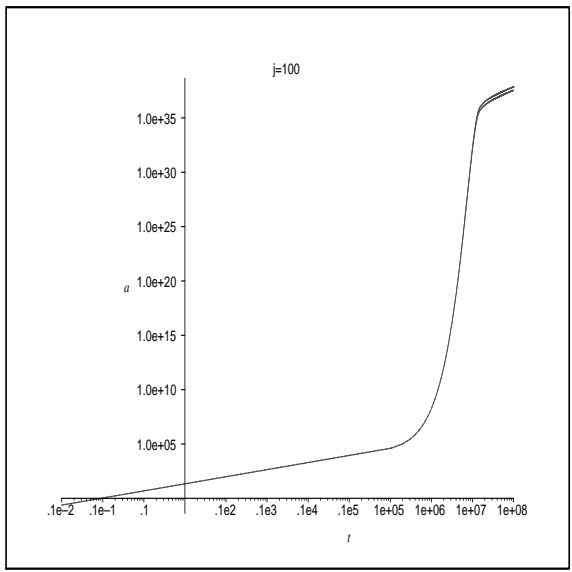}
\end{center}
\vskip-0.5cm \caption{Time dependence of the scale factor for
$j~=~100$. Dashed and solid curves correspond to the standard and
Plebanski models respectively} \label{fig2a}
\end{figure}

\begin{figure}
\begin{center}
\includegraphics[width=7.5cm,height=7.5cm, angle=270]{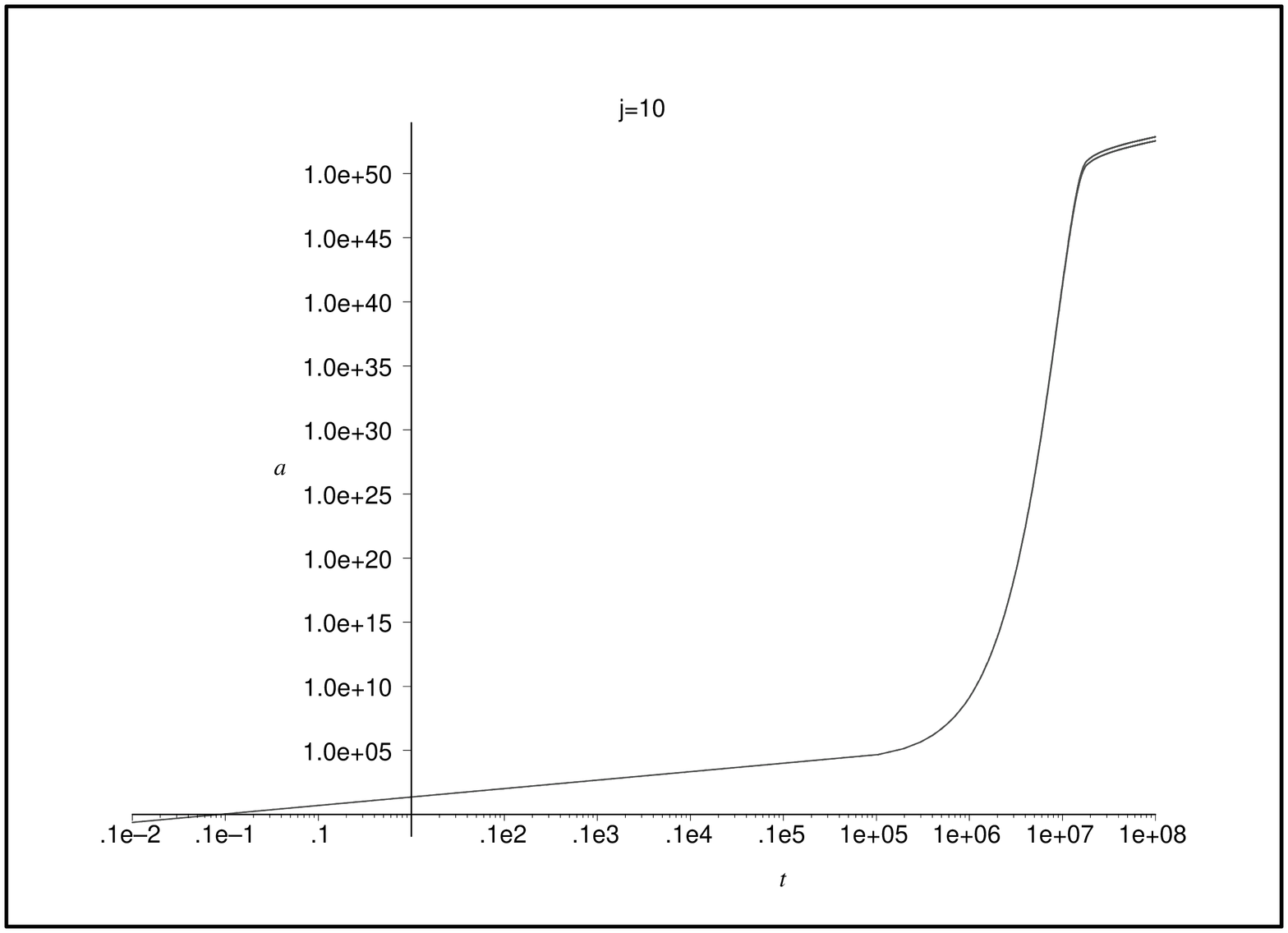}
\end{center}
\vskip-0.5cm \caption{Time dependence of the scale factor for
$j~=~10$. Dashed and solid curves correspond to the standard and
Plebanski models respectively} \label{fig2b}
\end{figure}

\begin{figure}
\begin{center}
\includegraphics[width=7.5cm,height=7.5cm,angle=270]{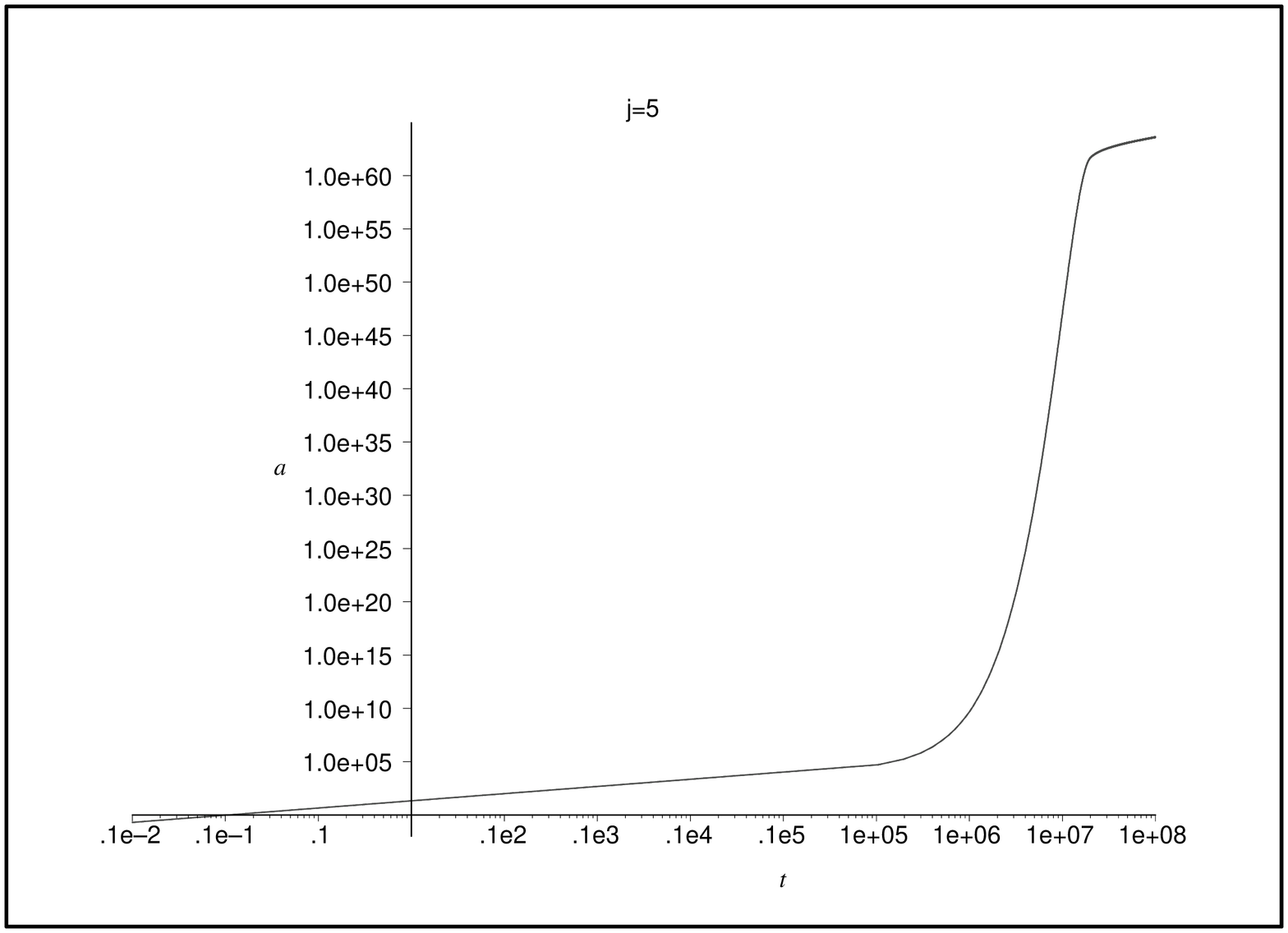}
\end{center}
\vskip-0.5cm \caption{Time dependence of the scale factor for
$j~=~5$. Dashed and solid curves correspond to the standard and
Plebanski models respectively} \label{fig2c}
\end{figure}

The three graphs below depict the time dependence of the the
scalar field, $\phi(t)$ (Fig. \ref{fig1a}-\ref{fig1c}), and the
scale factor, $a(t)$ (Fig. \ref{fig2a}-\ref{fig2c}), for different
values of the parameter $j$. Specifically we chose $j~=~100,10$
and 5. The qualitative dynamics for all the figures can be
described as follows. Initially: $a\ll a_*, \,D(q)\ll~1,\,{\cal D}
>~6$. Thus the right hand side of the Raychaudhuri equations (\ref{RaySC} ,\ref{RaySC_Pl}) is
positive (for the potential term in (\ref{RaySC_Pl}) can be
neglected), the inflaton goes up the potential hill driving
super-inflation ($\dot H>0$). When ${\cal D}\approx 6$, the right
hand side of (\ref{RaySC}) and (\ref{RaySC_Pl}) becomes negative
thus ending the super-inflationary phase. The scalar field
proceeds up-hill until ${\cal D}\approx 3$, turning into the
slow-roll regime with the oscillatory behaviour.

As the first three figures suggest, the maximum magnitude of the
inflaton, $\phi_{max}$, becomes greater for smaller $j$s. Further
numerical analysis shows that $\phi_{max}$ exceeds the required
value of 3M$_{P}$ at $j \approx 5$. As was expected the Plebanski
model gives a bit lower inflation.

We also see that in the semi-classical regime the evolution of the
inflaton of the Plebanski action differs from that of the standard
one. The difference is more significant for larger $j$s. Note,
however, that as $j$ decreases the discrepancy between the curves
describing the scalar field for the two models becomes more and
more negligible. In order to analyze this feature, let us first
rescale the variables $p$ and $p_\phi$. Choosing the units for the
scale factor to be $a_* \equiv \sqrt{\alpha} l_{\rm P}$ instead of
$l_{\rm P}$, along with $[p_\phi] \rightarrow [p_\phi]
\alpha^{3/2}$, we can completely get rid of $j$ in the equations
of motion. It now appears in the initial conditions bringing
modifications into the uncertainty principle (\ref{UncPr})
\be \Delta \phi \, \Delta p_\phi =\alpha^{3/2} \nonumber \ee
\begin{figure}
\begin{center}
\includegraphics[width=7.5cm,height=7.5cm, angle=270]{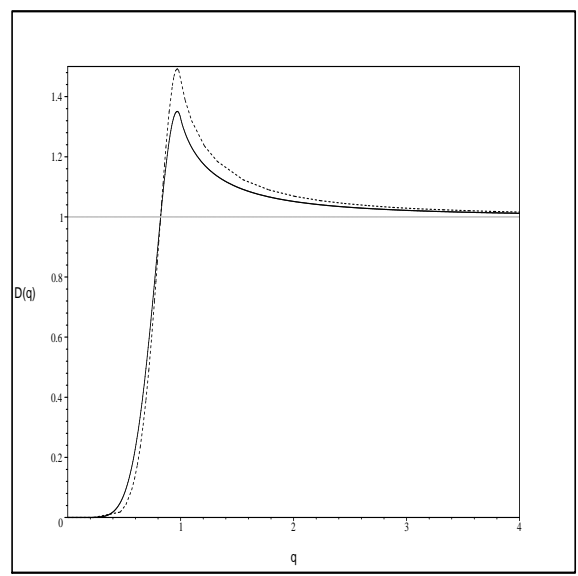}
\end{center} \vskip-0.5cm \caption{Spectrum of the geometrical
density operator $D(q)$ (solid curve) compared to $D(q)^{4/3}$
(dashed curve)} \label{fig3}
\end{figure}
Nonetheless, the initial conditions for both models still coincide,
and the only difference between the two sets of equations of motion
is the exponent of the geometrical density spectrum. In the graph
below we plot the dependencies of $D(q)$ and $D(q)^{4/3}$ (Fig.
\ref{fig3}). With the units chosen above, $a=a_*$ corresponds to
$q=1$.

Let us now consider the logarithmic derivative, $\cal D$. From the
quantum corrected Friedmann equations we see that at ${\cal D}=6$
the phase of superinflation ends, whereas at ${\cal D}=3$ the
anti-friction term of the Klein-Gordon equation turns into normal
friction. These values of the logarithmic derivative occur at
$q=0.91$ and $q=0.84$ respectively. Emphasize that they are the same
for both models and are in the range, where $D(q) \approx 1$ thus
the difference between $D$ and $D^{4/3}$ in Fig. \ref{fig3} is
small.

With the choice of units above, the initial value of the scale
factor is
\be a_0=\sqrt{3/j}. \ee
The latter implies that for greater values of $j$ the inflation
starts at smaller scales, when $D$ and $D^{4/3}$ may differ by
orders of magnitude. On the contrary, lower $j$ shifts the initial
scale factor towards the boundary of the semi-classical region
(for $j \le 3$ semi-classical regime is not existent at all).
Therefore when the parameter $j$ gets small, the difference
between $D$ and $D^{4/3}$, and hence between the standard and
Plebanski models, becomes insignificant.

\section{Discussion}
Summing up we would like to point out the main results of the
work.

We considered an alternative model for the gravitational action,
based on Plebanski's formulation. Classically both actions yield
the same physics. Especially this became manifest for the symmetry
reduced case, for the new action could be thought of as the
standard one with a specific lapse function. We should emphasize
again that the two models cannot be simply related by a choice of
lapse in the context of the full theory (non-homogeneous and
non-isotropic).

One can also start directly with the symmetry reduced
Einstein-Hilbert action, keeping the lapse freedom. Then the
choice of lapse, corresponding to the Plebanski model, arises
naturally. Indeed, if one sets the lapse function according to
(\ref{lapse}), one obtains the Hamiltonian constraint (\ref{HPl}),
which is manifestly self-adjoint and needs no symmetrization. Note
that such a choice of the lapse is unique.

The quantization in the framework of Einstein-Hilbert and
Plebanski formulations leads to different `evolution' equations in
both deep Planckian regime and semi-classical regime (i.e.
difference and differential equations respectively). We studied
the semi-classical approximation. The purpose of this work was
twofold. Firstly, we were to show that the Plebanski's action
would lead to semi-classical inflation and provide conditions for
successful slow-roll inflation, i.e. the value of the inflaton,
when entering the classical phase, would be $3M_{\rm P}$ or
greater. The second goal was to compare the phenomenological
implications of the two models, thus investigating the robustness
of the theory with respect to the lapse `ambiguity'.

Having solved the quantum corrected equations of motion we can
conclude the following. Common for both models, lower $j$s lead to
higher maximum values of the scalar field, attained during the
semi-classical phase of inflation. In order for the inflaton to
reach the value of $3M_{\rm P}$, the parameter $j$ has to be
approximately less than 5. Note, that if the parameter $j < 3$,
then there is no semi-classical regime at all. The inflaton,
however, reaches a higher maximum value, since the initial kinetic
energy is greater as $j$ decreases, provided fixed initial
conditions.

At the first glance, the correspondence between the parameter $j$
and the field maximum seems to disagree with the result of Ref.
\cite{CMB}. Specifically, the authors obtained the opposite
tendency: greater $j$ implied greater inflation. In fact, there is
no contradiction, and the discrepancy arises due to a different
approach to the choice of initial conditions. Consider, for
simplicity, the matter part of the standard Hamiltonian constraint
(\ref{Ham_init}). $j$ enters the expression through the kinetic
term
\be \label{Kin}
D(q) \frac{p_\phi^2}{2p^{3/2}} \equiv D^{-1}(q) \frac{\dot
\phi^2}{2p^{3/2}}
\ee
In \cite{CMB} the initial data is $\dot \phi$, whereas in this
paper, we have fixed $p_\phi$. Greater $j$ yields a smaller
initial scale factor $q_0 \equiv \left( a_0/a_*\right)^2=3/j$. The
latter corresponds to smaller $D(q_0)$ in (\ref{Kin}). Therefore,
if one keeps $\dot \phi_i$ the same and increases $j$, one gets an
increasing kinetic term, which leads to a higher maximum values of
the inflaton. On the contrary, fixing $p_{\phi i}$ and changing
$j$ implies a smaller initial kinetic energy and smaller
inflation. Similar statements hold true for the Plebanski model.

The comparison of the solutions to the quantum corrected equations
of motion, deriving from the alternative and the standard actions,
indicated that Plebanski's model always gives lower values for the
field maximum. Although the semi-classical dynamics is notably
different, when $j$ is large, the physically interesting solutions
($j<5$) agree very well for both models. Moreover, in the case of
large $j$, the solutions become very close to each other at large
scales, in spite of the significant difference between the two
models in the semi-classical regime. These results demonstrate
that the classical (observable) implications of the homogeneous
and isotropic LQC are indeed robust under the lapse freedom of the
theory.

{\bf Acknowledgments:} The author is grateful to Parampreet Singh
for detailed discussions, critical reading of the paper and
helpful comments. Many thanks should also go to M.~Bojowald and
K.~Vandersloot for thorough reading of the paper and valuable
suggestions, and T.~Pawlowski and E.~Bentivegna for comments. The
work was supported by the Duncan Fellowship of the Pennsylvania
State University.

\end{document}